\documentclass[journal]{IEEEtran}
\usepackage{amsmath,amssymb,verbatim,cite,citesort,amsopn,graphicx,citesort,url,color}
\usepackage{algorithm}
\usepackage{algorithmic}
\usepackage{multicol}
\usepackage{epsfig}
\usepackage{epstopdf}
\usepackage{subcaption}
\usepackage{balance}
\usepackage{enumerate}
\usepackage[font={small}]{caption}

\def\tablename{Table}
\renewcommand{\thetable}{\arabic{table}}

\newcommand{\pout}{p_{\text{out}}}

\newcommand{\pfa}{P_\text{{FA}}}
\newcommand{\pmd}{P_\text{{MD}}}
\newcommand{\up}{\text{{up}}}

\newcommand{\erf}{\text{erf}}

\newcommand{\xia}{\bar{\xi}}

\newcommand{\dB}{\text{{dB}}}

\newcommand{\LN}{\text{LN}}
\newcommand{\LU}{\text{LU}}

\newcommand{\by}{\mathbf{y}}

\newcommand{\PP}{\mathbb{P}}

\newtheorem{Proposition}{Proposition}

\newtheorem{Remark}{Remark}

\begin{document}

\title{On Covert Communication with Noise Uncertainty}

\author{
Biao~He,~\IEEEmembership{Member,~IEEE,}
Shihao~Yan,~\IEEEmembership{Member,~IEEE,}
Xiangyun~Zhou,~\IEEEmembership{Member,~IEEE,}
and Vincent~K.~N.~Lau,~\IEEEmembership{Fellow,~IEEE}\vspace{-8mm}\\
\thanks{B. He and V. K. N. Lau are with the Department of Electronic and Computer Engineering,
The Hong Kong University of Science and Technology, Hong Kong (email:
\{eebiaohe, eeknlau\}@ust.hk).}
\thanks{S. Yan and X. Zhou are with the Research School of Engineering, The Australian National University, ACT 2601, Australia (e-mail: \{shihao.yan, xiangyun.zhou\}@anu.edu.au).}
}


\maketitle

\begin{abstract}
Prior studies on covert communication with noise uncertainty adopted a worst-case approach from the warden's perspective.
That is, the worst-case detection performance of the warden is used to assess covertness, which is overly optimistic.
Instead of simply considering the worst limit, in this work, we take the distribution of noise uncertainty into account to evaluate the overall covertness in a statistical sense. Specifically, we define new metrics for measuring the covertness, which are then adopted to analyze the maximum achievable rate for a given covertness requirement under both bounded and unbounded noise uncertainty models.
\end{abstract}

\begin{IEEEkeywords}
Physical layer security, covert communication, noise uncertainty.\vspace{-3.5mm}
\end{IEEEkeywords}

\section{Introduction}\label{sec:Intro}

Secure communication against eavesdropping has drawn unprecedented attention from wireless researchers~\cite{Zhou_13_Physical} due to a fast growing amount of private and sensitive data  transmitted over wireless media.
In some situations, preventing eavesdropping is not enough and one wants to transmit covertly. For example, one may wish to communicate covertly to keep the privacy or avoid possible attacks. For some military operations,  an army may want to communicate without being detected by the enemy in order to hide its presence or movement.


Considering additive white Gaussian noise (AWGN) channels, a square root law was found in~\cite{Bash_13_limits}. That is, with an arbitrarily low probability of being detected, one can reliably send at most $\mathcal{O}\left(\sqrt{n}\right)$ bits over $n$ uses of the channel,
which indicates that the asymptotic covert rate approaches zero.
Lee \textit{et al.}~\cite{Lee_15_AchievingUC} found that the square root law can be broken when the warden has uncertainty about its receiver's noise power, and the asymptotic covert rate approaches a non-zero constant.
In~\cite{Lee_15_AchievingUC}, the detection performance at the warden was analyzed by a worst-case approach from the warden's perspective, where the actual noise power is assumed to be at the worst-case scenario for the detection.
Such an approach to analyzing the detection performance with noise uncertainty was
originally adopted in~\cite{Tandra_08_SNR} for the study of spectrum sensing, where one wanted to ensure that secondary users can always detect the communication between primary users, even in the worst-case scenario.
However, this approach has an important limitation for evaluating the detection performance at the warden for covert communication, since any true level of covertness cannot be guaranteed by assuming a worst-case scenario from the warden's perspective.

In this work, we restudy covert communication with the consideration of noise uncertainty.
To address the limitation of the existing approach, we propose new approaches for evaluating the covertness of the systems with noise uncertainty.
We analyze the systems with both bounded and unbounded models of the noise uncertainty, which are distinguished by whether the uncertainty range is finite. Our analysis takes the distribution of noise uncertainty into account, and hence, gives new results on covertness significantly different from the existing analysis which simply considered the worst limit of noise uncertainty. We show that our approaches are appropriate to study the covertness of the system with either bounded or unbounded noisy uncertainty, while the existing approach overestimates the covertness of the system.




\section{System Model}
Consider the scenario where Alice attempts to transmit messages to Bob  with a low probability of being detected by Willie. We consider the AWGN channel with real-valued signals.\footnote{Although real-valued signals are considered for ease of exposition, the analysis  can be easily extended to complex signals.}
Alice, Bob, and Willie each have a single antenna.
The normalized transmitted signals from Alice are denoted by $x[n]\sim\mathcal{N}\left(0,1\right)$, where $n=1, \cdots, N$.
When transmission happens, the received signal at Bob or Willie is given by
\vspace{-1mm}
\begin{equation}\label{}
  y_i[n]=\sqrt{\frac{P_t}{r_i^\alpha}}x[n]+v_i[n], \quad i=b~\text{or}~w,
\end{equation}
where $b$ and $w$ denote Bob and Willie, respectively, $P_t$ denotes the transmit power,  $v_i[n]\sim\mathcal{N}\left(0,\sigma_i^2\right)$   denote the noise at the receiver, $r_i$ denotes the distance from the transmitter to the receiver, and $\alpha$ denotes the path loss exponent. 
We assume that $x[n], v_b[n],$ and $v_w[n]$ are independent of each other. 


\subsection{Hypothesis Testing Problem at Willie}\label{sec:SM_HT}
Willie attempts to distinguish the following two hypotheses:\\
$\mathcal{H}_0:y_w[n]=v_w[n]$ and
$\mathcal{H}_1:y_w[n]=\sqrt{P_t/r_w^\alpha}x[n]+v_w[n]$.\\
Based on the received vector $\by_w\!=\!\left(y_w[1],\cdots,y_w[n]\right)$,
Willie makes a binary decision on whether the received signal is noise or signal plus noise.
We assume that Willie adopts a radiometer as the detector~\cite{Lee_15_AchievingUC}.
The test statistic is given by $T(\by_w)=(1/N)\sum_{n=1}^{N}\left|y_w[n]\right|^2$.
Denote $D_0$ and $D_1$ as the decisions that the received signal is noise and that the received signal is signal plus noise, respectively.
The false alarm probability and the misdetection probability are defined as
$ \pfa=\PP\left(D_1\mid\mathcal{H}_0\right)$
and
$  \pmd=\PP\left(D_0\mid\mathcal{H}_1\right)$,
respectively, where $\PP$ denotes the probability measure.
Willie wishes to minimize
$
  \xi=\pfa+\pmd,
$
and the ultimate objective of covert communication is to guarantee
$
  \xi=\pfa+\pmd\ge1-\epsilon,
$
for a positive and arbitrarily small $\epsilon$.  
In this paper, we focus on the case of $N\rightarrow\infty$~\cite{Lee_15_AchievingUC}, which allows Willie to observe an infinite number of samples.
Denote the average received signal power at Willie as $P_w=P_t/r_w^\alpha$.
We assume that $P_w$ is known.
Without the noise uncertainty, we have the following approximations for the case of large $N$ from the central limit theorem~\cite{Durrett_04_bookProb,Tandra_08_SNR}:
\begin{small}
$
T(\by_w)|\mathcal{H}_0\sim\mathcal{N}\left(\sigma_w^2,2\sigma_w^4/N\right),
T(\by_w)|\mathcal{H}_1\sim\mathcal{N}\left(P_w\!+\!\sigma_w^2,2\left(P_w+\sigma_w^2\right)^2/N\right),
\pfa\!\approx\!Q\!\left(\gamma\!-\!\sigma^2_w\left(\sqrt{2/N}\sigma^2_w\right)\right),
$\\
$
\pmd\!\approx\! 1\!-\!Q\!\left(\!\gamma-\left(P_w+\sigma^2_w\right)/\left(\sqrt{2/N}\left(P_w+\sigma^2_w\right)\right)\!\right),
$
\end{small}
where $\gamma$ is the threshold of Willie's detector and $Q(\cdot)$ denotes the tail probability of the standard normal distribution. 

Based on the above approximations, we have, as $N\rightarrow\infty$,
\begin{equation}\label{eq:pc01}
  \xi\left(\sigma_w^2,\gamma\right)\approx\left\{\begin{array}{ll}
  0\;, &\mbox{if}~   \sigma_w^2\le\gamma\le P_w+\sigma_w^2\\
  1\;, &\mbox{otherwise.}
\end{array}\right.
\end{equation}
If $\sigma_w^2$ is known without uncertainty, Willie can simply set any $\gamma\in\left[\sigma_w^2, P_w+\sigma_w^2\right]$ to ensure $\xi\rightarrow0$, which implies that  Willie can detect the communication without any error.

\subsection{Noise Uncertainty}\label{sec:noiseUncertainty}

The lack of knowledge of the exact noise power is called noise uncertainty~\cite{Shellhammer_06_stdPerformance}. In practice, the sources of background noise include thermal noise, quantization noise, imperfect filters, ambient wireless signals, etc. Noise uncertainty is almost unavoidable due to, e.g., temperature change, environmental noise change, and calibration error. Therefore, the consideration of noise uncertainty is practical and necessary for the study related to power detection.
Furthermore, as pointed out in~\cite{Lee_15_AchievingUC} and~\cite{Sobers_16_CovertJammer}, legitimate users can also intentionally generate interference signals to increase and ensure the noise uncertainty at Willie for the purpose of covert communication.


In this work, we consider two models of the noise uncertainty at Willie, which are detailed as follows. 


\subsubsection{Bounded Uncertainty Model}
For the bounded uncertainty model, the exact noise power $\sigma^2_w$ lies in a finite range around the nominal noise power. Denoting the nominal noise power as $\sigma^2_{n}$, we assume that $\sigma^2_{w,\dB}\in[\sigma^2_{n, \dB}-\rho_{\dB}, \sigma^2_{n, \dB}+\rho_{\dB}]$ in the dB domain~\cite{Shellhammer_06_stdPerformance,Tandra_08_SNR}, where
$\sigma^2_{w,\dB}=10\log_{10}{\sigma^2_{w}}$, $\sigma^2_{n, \dB}=10\log_{10}{\sigma^2_{n}}$, and $\rho_{\dB}=10\log_{10}(\rho)$ is the parameter that  quantifies the size of the uncertainty.
We further assume that $\sigma^2_{w,\dB}$ is uniformly distributed in its uncertainty range in the dB domain~\cite{Shellhammer_06_stdPerformance,Zeng_09_Reliability}. 
Then, the log-uniform distribution of $\sigma^2_w$ for the bounded uncertainty model is given by
\begin{equation}\label{eq:apriorNPdis}
  f_{\sigma_w^2}(x)=\left\{\begin{array}{ll}
  \frac{1}{2\ln\left(\rho\right)x}
  \;, &\mbox{if}~\frac{1}{\rho}\sigma^2_{n}\le\sigma_w^2\le\rho\sigma^2_{n}\\
  0
  \;, &\mbox{otherwise.}
  \end{array}\right.
\end{equation}


\subsubsection{Unbounded Uncertainty Model}
For the unbounded uncertainty model, the exact noise power $\sigma^2_w$ does not necessarily lie in a finite range. Instead, $\sigma^2_{w,\dB}\in[-\infty, +\infty]$.
We assume that the difference between the exact noise power and the nominal noise power in the dB domain follows a normal distribution~\cite{Sonnenschein_92_radiometric,Jouini_11_energydlu},
i.e., $\Delta=\sigma^2_{w,\dB}-\sigma^2_{n, \dB}\sim\mathcal{N}\left(0, \sigma_{\Delta,\dB}^2\right)$.
Denoting $k=\ln(10)/10$, the log-normal distribution of $\sigma^2_w$ for the unbounded uncertainty model is given by
\begin{align}\label{eq:fsigmaLNo}
  f_{\!\sigma_w^2}\!(\!x\!)
 \!=\!\left\{\!\!\!\!\begin{array}{ll}
  \frac{1}{x\!\sqrt{2\pi k^2\sigma_{\!\Delta,\dB}^2}}\exp\!\left(\!\!-\frac{\left(\ln(\!x\!)-k\sigma^2_{\!n, \dB}\right)^{\!2}}{2k^2\sigma_{\!\Delta,\dB}^2}\!\right)
  \!, \!\!\!\!&\mbox{if}~x>0\\
  0
  \;, &\mbox{otherwise.}
  \end{array}\right.\!
 \end{align}

We assume that the statistics of noise uncertainty, i.e., \eqref{eq:apriorNPdis} or \eqref{eq:fsigmaLNo}, are known, while the exact noise power is unknown. The noise uncertainty at Bob is not considered in this work, since it does not affect the performance of the covertness.


\section{Approach to Measuring Covertness}
As introduced in Section~\ref{sec:SM_HT}, the covertness is examined by $\xi=\pfa+\pmd$. For a known noise power $\sigma_w^2$ and detection threshold $\gamma$, $\xi\left(\sigma_w^2,\gamma\right)$ is given by~\eqref{eq:pc01}.
In what follows, we discuss the approach to analyzing the covertness with the consideration of noise uncertainty. We first point out a severe limitation of the existing approach adopted in~\cite{Lee_15_AchievingUC}, and then introduce the newly-adopted approaches in this work. 

\subsection{Limitation of Existing Approach}
In the existing work~\cite{Lee_15_AchievingUC}, a robust statistics approach was adopted to measure covertness with noise uncertainty.
Specifically, the upper limit of $\xi$ with noise uncertainty is adopted as the measure and is given by
\begin{equation}\label{eq:xiorg_robust}
  \xi_{\up}=\min_{\gamma}\max_{\sigma_w^2} \xi\left(\sigma_w^2,\gamma\right).
\end{equation}
The requirement of covert communications is $\xi_{\up}\ge1-\epsilon$.
Note that \eqref{eq:xiorg_robust} characterizes the worst-case performance from Willie's perspective.
The robust statistics approach was originally adopted in~\cite{Tandra_08_SNR} for the study of spectrum sensing with noise uncertainty where the aim is to ensure that users can always detect communications, even in the worst-case scenario.
In contrast, for the purpose of covert communication, one wants to ensure that Willies cannot detect the communication. The robust statistics approach from Willie's perspective is actually the most idealized (non-robust) approach to achieve covert communication from the legitimate user's perspective.


A system that achieves $\xi_{\up}>1-\epsilon$ cannot, in fact, guarantee
any true level of covertness.
For example, it is found in~\cite{Lee_15_AchievingUC} that a system with
\begin{equation}\label{eq:P_th_robust_wrong}
  P_w<\left(\rho-\frac{1}{\rho}\right)\sigma_{n}^2
\end{equation}
can ensure that $\xi_{\up}\ge1-\epsilon$ for an arbitrarily small $\epsilon$, when the bounded uncertainty model of $\sigma_w\in\left[(1/\rho)\sigma_n^2,\rho\sigma_n^2\right]$ is adopted. However, a system with
$P_w\rightarrow\left(\rho-1/\rho\right)\sigma_{n}^2$ will result in $\min_{\gamma}\xi\left(\gamma,\sigma_w^2\right)<1-\epsilon$ as long as the exact noise power at Willie is not at the worst-case limit.


%
%
%

\subsection{Newly-Adopted Approach}
In this work, we adopt two new approaches to measure the covertness with noise uncertainty.
Instead of focusing on the worst-case scenario at Willie, we analyze the overall performance at Willie with the noise uncertainty. The two approaches are described as follows.



\subsubsection{Bayesian Statistics}
The first approach is the Bayesian statistics approach~\cite{Shellhammer_06_stdPerformance}.
Specifically, the average of $\xi$ over the a priori distribution of the noise power is adopted as the measure, which is given by
\begin{equation}\label{eq:xiagamma}
\xia=\min_{\gamma}\int_{0}^{\infty}\xi\left(\sigma_w^2,\gamma\right)f_{\sigma_w^2}\left(\sigma_{w}^2\right)\mathrm{d}\sigma_w^2.
\end{equation}
We name $\xia$ as the average covert probability, and the requirement of covert communication is given by $\xia\ge1-\epsilon$.
The average covert probability captures the average covertness performance of the system over multiple times of communications/experiments.




\subsubsection{Outage-Based Approach}
The second approach is the outage-based approach.
Specifically, the probability that $\xi<1-\epsilon$ is adopted as the measure, which is given by
\begin{equation}\label{eq:poutgamma}
   \pout=\min_{\gamma}\PP\left(\xi\left(\sigma_w^2,\gamma\right)<1-\epsilon\right). 
   \vspace{-2mm}
\end{equation}
We name $\pout$ as the covert outage probability.
The requirement of covert communication is given by $\pout\le\delta$, where $\delta$ denotes the maximum acceptable outage probability.
The covert outage probability captures the probability that covert communication fails in such scenarios. Note that the concept of covert outage probability is similar to the concept of secrecy outage probability~\cite{Zhou_11} which characterizes the probability that secure communication fails.


\begin{Remark}\label{remark:2}
Both $\xia$ and $\pout$ depend on the a priori distribution of the noise power. Therefore, the covertness depends on the uncertainty model of the noise power.
Mathematically, $\pout=1-\xia$ in this work with the consideration of $N\!\rightarrow\!\infty$, and the  requirement of $\xia\ge1-a$ is equivalent to the requirement of $\pout\le a$.
This is because  $\xi\left(\sigma_w^2,\gamma\right)$ in~\eqref{eq:pc01} is equal to either $0$ or $1$. However, the expressions for $\xia$ and $\pout$ will be different when $\xi\left(\sigma_w^2,\gamma\right)$ starts taking values within the range of $(0,1)$, which happens in  scenarios where Willie has a finite number of samples~$N$~\cite{Lee_14_con_achieving_siso}.
\end{Remark}




%

\vspace{-3mm}
\section{Covert Rate}
Now, we investigate the maximum rate at which Alice can reliably communicate with Bob subject to a constraint on the average covert probability, namely, the covert rate.
Denoting the covert rate as $R$, the problem is formulated as
\begin{subequations}\label{}
\begin{align}\label{}
  \max_{P_t} &~~~ R=\frac{1}{2}\log_2(1+\frac{P_t}{r_b^\alpha\sigma_b^2}) \\
 \mathrm{s.t.} &~~~~ \xia\ge1-\epsilon.
\end{align}
\end{subequations}
As explained in Remark~\ref{remark:2}, the requirement of $\xia\!\ge\!1\!-\!a$ is equivalent to the requirement of $\pout\le a$ in this work. Thus, we do not consider the requirement of $\pout\le\delta$ in this section. 

In the following, we derive the covert rate in systems with different models of noise uncertainty. For each model, we first obtain the received power threshold at Willie below which $\xia\ge1-\epsilon$, and then present the corresponding covert rate.



\vspace{-2mm}
\subsection{Bounded Uncertainty Model:}
The noise power at Willie is assumed to follow a log-uniform distribution given by~\eqref{eq:apriorNPdis}.
\begin{Proposition}\label{Theo:PthBayes}
   With the log-uniformly distributed noise power, the received power threshold at Willie below which $\xia\ge1-\epsilon$ is given by
\begin{equation}\label{eq:P_th_ave_r}
  P_{\LU}=\left(\rho^{2\epsilon-1}-\frac{1}{\rho}\right)\sigma_{n}^2.
\end{equation}
\end{Proposition}
\begin{IEEEproof}  The detailed proof is omitted here due to space limitation. Instead, the important steps in the proof are given as follows:

If $P_w\ge\left(\rho-(1/\rho)\right)\sigma_{n}^2$, Willie can set $\gamma=\rho\sigma_{n}^2$ to have $\xia\rightarrow0$, so that he can detect the communication without any error. 
We then focus on the case that $P_w<\left(\rho-(1/\rho)\right)\sigma_{n}^2$.
One can find that Willie's optimal detection threshold is
\begin{align}\label{eq:optgamma1}
  \!\gamma^*\!\!=\!\arg\min_{\gamma}\!\!\int_{\frac{1}{\rho}\sigma_{n}^2}^{\rho\sigma_{n}^2}\!\!\!\xi\!\left(\sigma_w^2,\gamma\right)\!f_{\!\sigma_w^2}\!\left(\sigma_{w}^2\right)\mathrm{d}\sigma_w^2   \!=\!P_w\!+\!(\!1/\rho)\sigma_{n}^2.\!\!
\end{align}
With~\eqref{eq:optgamma1}, we have
\begin{align}\label{}
  \!\xia\!=\!\int_{\frac{1}{\rho}\sigma_{n}^2}^{\rho\sigma_{n}^2}\!\!\xi\!\left(\!\sigma_w^2,\!\gamma^*\!\right)\!f_{\sigma_w^2}\!\left(\!\sigma_{w}^2\!\right)\!\mathrm{d}\sigma_w^2 \!=\!\frac{1}{2\!\ln(\rho)}\!\ln\!\left(\!\frac{\rho^2\sigma_{n}^2}{\rho P_w\!+\!\sigma_{n}^2}\!\right)\!.
\end{align}
Solving the inequality $\xia\!\ge\!1\!-\!\epsilon$ for $P_w$ completes the proof. 
\end{IEEEproof}

\begin{Remark}\label{Remark:3}
We note from~\eqref{eq:P_th_ave_r} that $P_{\LU}\rightarrow0$ as $\epsilon\rightarrow0$, which indicates that a sufficiently small $P_w$ is required to ensure an arbitrarily small $\epsilon$.
We also note from~\eqref{eq:P_th_ave_r} that $P_{\LU}\rightarrow(\rho-1/\rho)\sigma_{n}^2$ as $\epsilon\rightarrow1$, which is equal to the threshold to ensure $\xi_{\up}\ge1-\epsilon$ with $\epsilon\rightarrow0$; see~\eqref{eq:P_th_robust_wrong}. Therefore, we confirm that a designed system based on the measure of $\xi_{\up}$ cannot guarantee any true level of covertness.
\end{Remark}

With~\eqref{eq:P_th_ave_r}, the transmit power is limited by $P_t=r_w^\alpha P_{\LU}$. The covert rate for systems with a log-uniformly distributed noise power at Willie is given by
\begin{align}\label{}
  R_{\LU}&=\frac{1}{2}\log_2\left(1+\frac{r_w^\alpha P_{\LU}}{r_b^\alpha\sigma_b^2}\right).
\end{align}



\vspace{-5mm}
\subsection{Unbounded Uncertainty Model}
The noise power at Willie is assumed to follow the log-normal distribution given by~\eqref{eq:fsigmaLNo}.
The closed-form expression for the power threshold is mathematically intractable due to the complicated expression of $f_{\sigma_w^2}(x)$ in~\eqref{eq:fsigmaLNo}. Instead, we give an approximated power threshold in the following proposition.
\begin{Proposition}\label{Theo:PthBayes_lognormal}
   With the log-normally distributed noise power, the received power threshold at Willie below which $\xia\ge1-\epsilon$ is approximated by
\begin{equation}\label{eq:PLN}
  P_{\LN}\!\approx\!\left\{\begin{array}{ll}
 \!\!\! 2\sqrt{2\phi_2}~\erf^{-1}\!\left(\phi_3\epsilon\right) \;\!\!, ~~~~~\mbox{if}~\epsilon<\frac{1}{\phi_3}\erf\left(\frac{\phi_1}{\sqrt{2\phi_2}}\right)\\
  \!\!\!\phi_{\!1}\!\!-\!\!\sqrt{2\phi_2}~\erf^{-\!1}\!\!\left(\!\erf\!\left(\!\frac{\phi_1}{\sqrt{2\phi_2}}\!\right)\!\!-\!\!2\phi_3\epsilon\!\right)\;\!\!, \mbox{otherwise,}
  \end{array}\right.
\end{equation}
where
$
 \!\!\phi_1\!\!=\!\!\exp\!\left(k\sigma^2_{\!n, \dB}\!+\!k^2\sigma_{\!\Delta, \dB}^2/2\right),
  \phi_2\!\!=\!\!\left(\exp\!\left(k^2\sigma_{\!\Delta, \dB}^2\!\right)\!-\!1\right)
$
$  \exp\left(2k\sigma^2_{n, \dB}+k^2\sigma_{\Delta, \dB}^2\right),
$
$
  \phi_3=\frac{1}{2}\left(1\!-\!\erf\left(-\frac{\phi_1}{\sqrt{2\phi_2}}\right)\right),
$
$\erf(\cdot)$ denotes the error function, and $\erf^{-1}\!\left(\cdot\right)$ denotes the inverse error function.
\end{Proposition}
\begin{IEEEproof}
The detailed proof is omitted here due to space limitation. Instead, the important steps in the proof are given as follows:

When the level of noise uncertainty is small, the expression for $f_{\sigma^2}\left(x\right)$ in~\eqref{eq:fsigmaLNo} can be approximated by a Gaussian function~\cite{Hammouda_12_NNICRS}, given by
\begin{equation}\label{eq:appfrac1x}
 f_{\sigma^2}\left(x\right)\approx
  \frac{1}{\sqrt{2\pi\phi_2}\phi_3}\exp\left(-\frac{\left(x-\phi_1\right)^2}{2\phi_2}\right),~x>0.
\end{equation}
With~\eqref{eq:appfrac1x}, one can find that Willie's optimal detection threshold is $\gamma^*=\max\left\{\phi_1+P_w/2,~P_w\right\}$. We then have
\begin{align}\label{}
   \xia\!
   =\!\left\{\begin{array}{ll}
   \!1-\frac{1}{\phi_3}\erf\left(\frac{P_w}{2\sqrt{2\phi_2}}\right)
   \;\!\!, &\mbox{if}~P_w\!<\!2\phi_1\\
   \!1\!-\! \frac{1}{2\phi_3}\!\left(\!\erf\!\left(\!\frac{\phi_1}{\sqrt{2\phi_2}}\right)\!-\!\erf\!\left(\!\frac{\phi_1-P_w}{\sqrt{2\phi_2}}\!\right)\!\right)
  \;\!\!, &\mbox{otherwise.}
  \end{array}\right.\!\!\!\!\!
\end{align}
Solving the inequality $\xia\!\ge\!1\!-\!\epsilon$ for $P_w$ completes the proof.
\end{IEEEproof}

\begin{Remark} We note from~\eqref{eq:PLN} that $P_{\LN}\rightarrow0$ as $\epsilon\rightarrow0$.
We also note from~\eqref{eq:PLN} that $P_{\LU}\rightarrow\infty$ as $\epsilon\rightarrow1$, which, in fact, would be the threshold to ensure $\xi_{\up}\ge1-\epsilon$ with $\epsilon\rightarrow0$ under the unbounded uncertainty model.
Intuitively, it is unreasonable that we could achieve  covert communication with an arbitrarily large power so long as Willie has an unbounded noise uncertainty model. Thus, we again confirm that $\xi_{\up}$ is not appropriate in evaluating the covertness of systems with noise uncertainty.
\end{Remark}




With~\eqref{eq:P_th_ave_r}, the transmit power is limited by $P_t=r_w^\alpha P_{\LN}$. The covert rate for systems with a log-normally distributed noise power at Willie is given by
\begin{align}\label{}
  R_{\LN}&=\frac{1}{2}\log_2\left(1+\frac{r_w^\alpha P_{\LN}}{r_b^\alpha\sigma_b^2}\right).
\end{align}

\section{Numerical Results}

\begin{figure}[!htb]
\centering
\vspace{0mm}
\includegraphics[height=1.8in,width=2.2in]{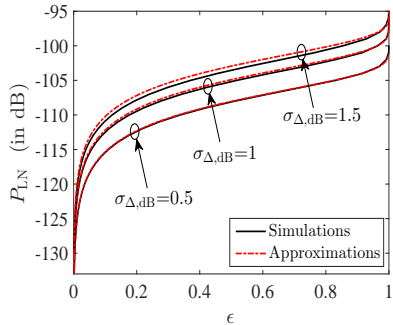}
\vspace{0mm}
\caption{Illustration of the accuracy of the approximated $P_{\LN}$.}
\vspace{0mm}  \label{fig:appcheck} 
\end{figure}

We first examine the accuracy of the obtained approximation of $P_{\LU}$ given in Proposition~2.
Figure~\ref{fig:appcheck} compares the approximated $P_{\LN}$ in~\eqref{eq:PLN} and the numerically obtained $P_{\LN}$ by Monte-Carlo simulations. The nominal noise power at Willie in the dB domain is $\sigma^2_{n,\dB}=-100$. As shown in the figure, the approximations match precisely with the simulations. The accuracy of the approximation improves as the variance of the noise decreases, and the approximation error is almost unnoticeable for the case of $\sigma_{\Delta,\dB}=0.5$.

\begin{figure}[t!]
    \centering
    \begin{subfigure}[t]{0.5\columnwidth}
        \centering
        \includegraphics[height=1.57in,width=1.85in]{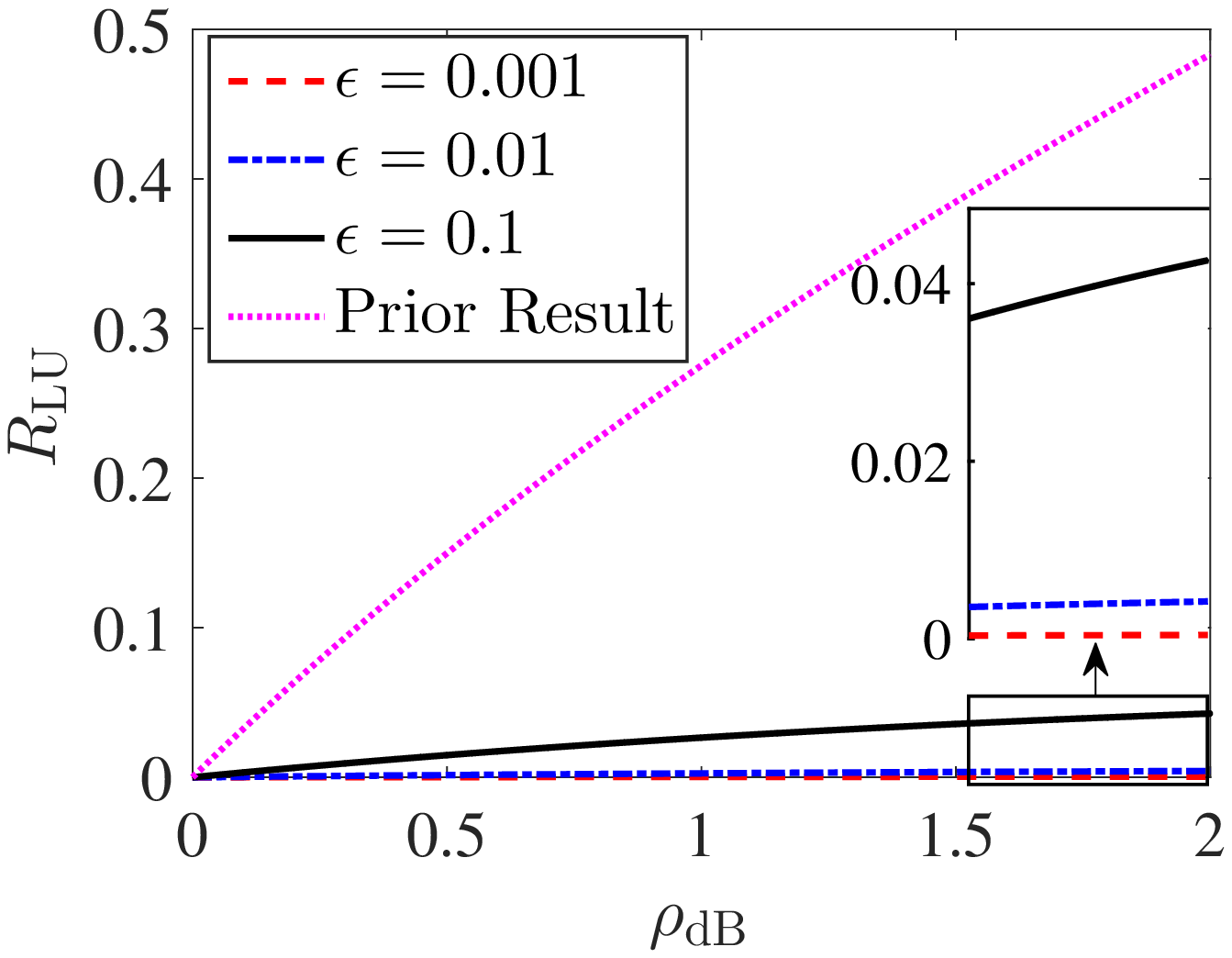}
        \caption{Bounded uncertainty model.} \label{fig:LU}\vspace{-1mm}
    \end{subfigure}%
    ~\begin{subfigure}[t]{0.5\columnwidth}
        \centering
        \includegraphics[height=1.57in,width=1.85in]{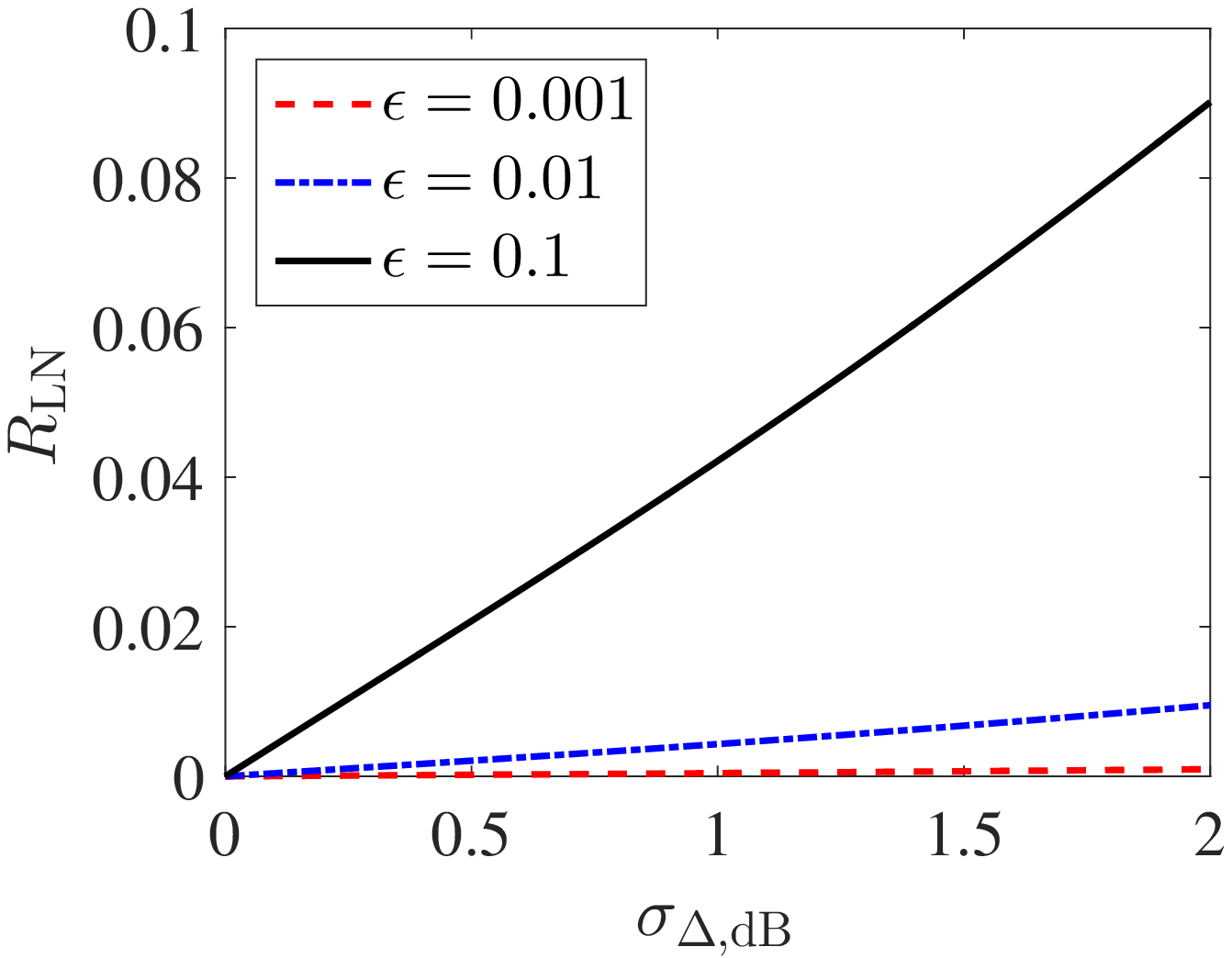}
        \caption{Unbounded uncertainty model.} \label{fig:LN}\vspace{-1mm}
    \end{subfigure}
    \caption{Covert rate versus noise uncertainty level.} \label{fig:LUN}\vspace{-5mm}
\end{figure}

We now show the covert rates of systems with different levels  of noise uncertainty at Willie. Figure~\ref{fig:LUN}(a)  plots $R_{\LU}$ versus $\rho_{\dB}$ for the bounded noise uncertainty model with log-uniformly distributed noise power at Willie, while Figure~\ref{fig:LUN}(b) plots $R_{\LN}$ versus $\sigma_{\Delta,\dB}$ for the unbounded noise uncertainty model with log-normally distributed noise power at Willie.
The parameters are set as  $r_b\!=\!r_w$ and $\sigma^2_{b,\dB}\!=\!\sigma^2_{n,\dB}\!=\!-100$,~where $\sigma^2_{b,\dB}\!=\!10\log_{10}{\sigma^2_{b}}$.
For comparison, the curve of ``prior result"  in Figure~\ref{fig:LUN}(a) represents the achievable covert rate with an arbitrarily small probability of being detected derived in the prior work~\cite{Lee_15_AchievingUC}, which adopted $\xi_{\up}$ to measure the covertness. The curve of ``prior result" is not shown in Figure~\ref{fig:LUN}(b), since the achievable covert rate  based on $\xi_{\up}$ always goes to infinity.
As depicted in the figures, the covert rate increases as Willie's noise uncertainty increases and/or the required $\epsilon$ increases. The covert rate approaches zero if Willie does not have uncertainty of his noise power, i.e., $\rho_{\dB}, \sigma_{\Delta,\dB}\rightarrow0$.
According to our analysis, the covert rate still approaches zero as $\epsilon$ approaches zero subject to any level of noise uncertainty. We note that the prior result is much larger than the achievable covert rate according to our analysis. Thus, the adoption of $\xi_{\up}$ overestimates the covertness, and is not  appropriate in the study of covert communication.
\vspace{-2mm}


%
%
%

\section{Conclusion}
We have revisited covert communication with the consideration of noise uncertainty. The  average covert probability and the covert outage probability have been proposed to measure covertness when the warden has noise uncertainty. We have considered both the bounded and unbounded models of noise uncertainty. For each model, the rate below which one can communicate subject to a given requirement of covertness has been derived. Our results show that a positive cover rate is achievable with a low probability of being detected, while the covert rate approaches zero as the probability of being detected approaches zero.
\vspace{-2mm}\vspace{-1mm}

\end{document}